\documentclass[aip,amsmath,amssymb,reprint,citeautoscript,longbibliography]%
{revtex4-2}

\usepackage{graphicx}
\usepackage[utf8]{inputenc}
\usepackage{color}
\definecolor{goodgreen}{rgb}{0.1,0.5,0}
\definecolor{goodred}{rgb}{0.7,0,0}
\usepackage[colorlinks,urlcolor=goodgreen,citecolor=blue,%
linkcolor=goodred]{hyperref}
\usepackage{upgreek}
\usepackage{bm}

\usepackage{newtxtext,newtxmath}

\newcommand{\un}[1]{{\ensuremath{\,\text{#1}}}}

\newcommand{\Qcav}{\ensuremath{Q_\text{c}}}

\begin{document}

\title{Stepwise fabrication and optimization of coplanar waveguide resonator hybrid devices}

\author{N. Kellner}
\author{N. Hüttner}
\affiliation{Institute for Experimental and Applied Physics, University of
Regensburg, Universitätsstr.\ 31, 93053 Regensburg, Germany}
\author{M. Will}
\author{P. Hakonen}
\affiliation{QTF Centre of Excellence, Department of Applied Physics, Aalto 
University School of Science, P.O. Box 15100, 00076 Aalto, Finland}
\author{A. K. Hüttel}
\email{andreas.huettel@ur.de}
\affiliation{Institute for Experimental and Applied Physics, University of
Regensburg, Universitätsstr.\ 31, 93053 Regensburg, Germany}
\affiliation{QTF Centre of Excellence, Department of Applied Physics, Aalto 
University School of Science, P.O. Box 15100, 00076 Aalto, Finland}

\begin{abstract}
From the background of microwave-optomechanical experiments involving carbon 
nanotubes, the optimization of superconducting coplanar waveguide resonator 
devices is discussed. Two devices, one with unmodified geometry compared to 
previous work and one integrating several improvements, are lithographically 
built up step by step. After each step, the low temperature GHz transmission 
properties are retested. This allows to identify the impact of the fabrication 
and the geometry modification on the device properties. In addition, simplified 
circuit geometries are modeled numerically, confirming the experimental results 
and providing further insights for optimization.  
\end{abstract}

\maketitle


\section{Introduction}

Research in the field of optomechanics\cite{rmp-aspelmeyer-2014} has over the 
past years covered a wide range of material systems and parameter ranges. In 
terms of frequencies, experiments reach from optics to microwave technology; 
mechanical systems can be single atoms or macroscopic mirrors. In a 
comparatively recent development, also carbon nanotubes have been integrated 
with dispersive microwave optomechanical circuits.\cite{optomechanics, 
modelingomit} Given their high mechanical quality factors at cryogenic 
temperatures\cite{highq, nnano-moser-2014} and their properties as prototypical 
single electron devices,\cite{rmp-laird-2015} this paves the way for novel 
combinations of quantum transport and optomechanical manipulation.

\begin{figure*}[t]
\includegraphics{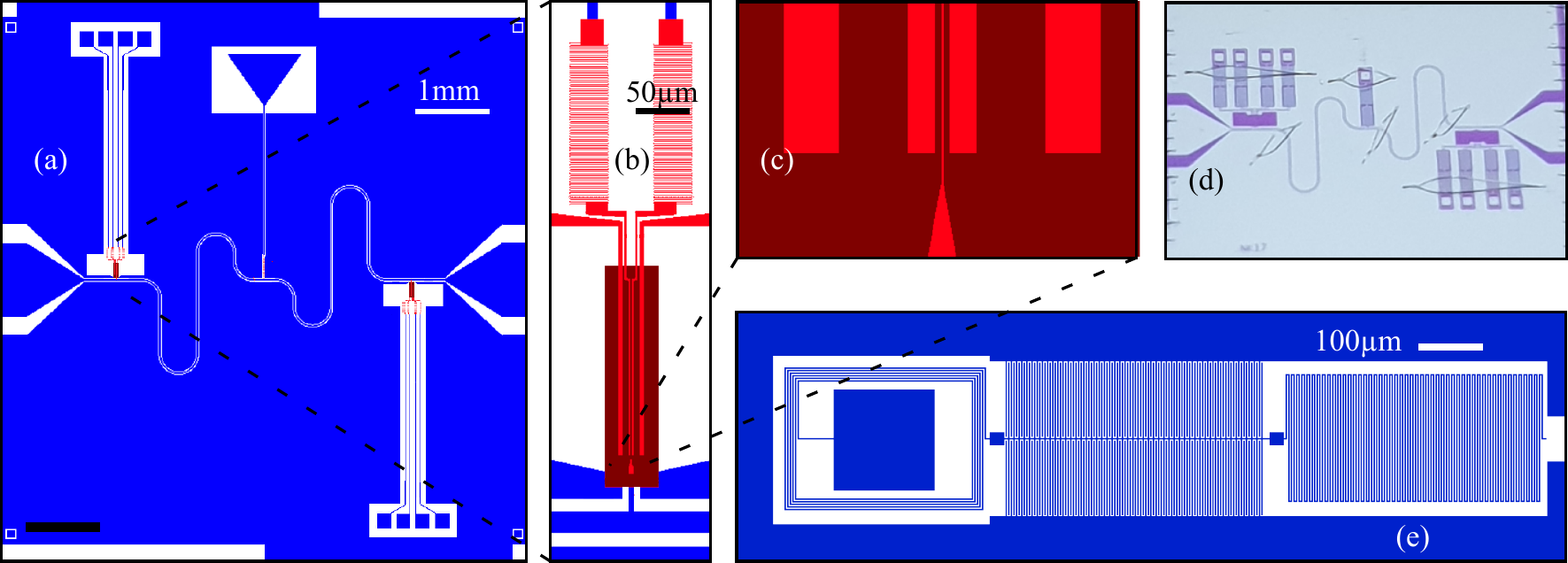} 
\caption{(a) Overall geometry of device A. Blue: Niobium layer, white: etched 
region with bare substrate surface, brown: gate insulator, red: gold electrodes 
and meanders. (b) Detail zoom of (a) in the carbon nanotube transfer 
region.\cite{forktransfer} (c) Detail zoom of (b), showing the gate finger 
between the source/drain contacts. Note that in the actual device the gate 
finger lies below, the contacts however above the gate insulator. (d) Overview 
photograph of device B after bonding. (e) Filter circuit as added to each of 
the dc connections in device B, consisting of a coil inductor, an interdigital 
capacitor, and a meander inductor.}
\label{figDevices}
\end{figure*}

The central limitation of the device measured in 
Refs.~\onlinecite{optomechanics, modelingomit}, with the geometry as also shown
in Fig.~\ref{figDevices}(a-c), was a very low quality factor $\Qcav \sim 500$
of the microwave resonator. While also fabrication defects may have played a
role there, with fluorinated resist flakes stuck to the central conductor of
the coplanar waveguide,\cite{optomechanics, modelingomit} the main challenge is
a more systematical one. Previous work in Regensburg has demonstrated standalone
high-\Qcav\ resonators.\cite{toscres} Inserting and contacting carbon nanotubes
into the device\cite{forktransfer} and performing both quantum transport and
microwave transmission measurements, however, requires the definition of dc
electrodes close to the coplanar waveguide resonator or even attaching to it.
These immediately lead to leakage of the GHz field.

Here, we present data on the fabrication and characterization of two niobium 
devices, A and B. Device A is geometrically close to the device of 
Refs.~\onlinecite{optomechanics, modelingomit}, while device B integrates larger
filters to block the signal leakage as well as further improvements. After each 
lithographic step the devices are cooled down and the resonator properties are 
tested at $T=4.2\un{K}$. This allows us to identify the fabrication steps 
detrimental to $Q_\text{c}$ as well as validate the effect of our optimizations.
Subsequently the transmission of simplified device geometries is modelled 
numerically, clearly confirming our experimental result and providing insight 
into the detailed mechanisms.

\section{Device geometry and fabrication steps, device A}

Figure~\ref{figDevices} displays the geometry of the two analyzed devices. In 
each case the substrate is a wafer of 500\,\textmu{}m compensation doped 
silicon covered by 500\,nm thermally grown SiO\textsubscript{2} and a 100\,nm 
niobium film. Fig.~\ref{figDevices}(a) shows the full lithograpy drawing of
device A, in its geometry identical to the device used in
Refs.~\onlinecite{optomechanics, modelingomit}; Fig.~\ref{figDevices}(b) and
Fig.~\ref{figDevices}(c) show details of it.

\begin{table}[b]
\begin{center}
\begin{tabular}{|c|p{6cm} |c|c|}
\hline
\multicolumn{2}{|l|}{\textbf{measured $\bm{Q_c}$ at $\bm{T=4.2\,\text{K}}$ }}
& \multicolumn{2}{c|}{\textbf{device}} \\
\multicolumn{2}{|l|}{\textbf{after fabrication step}} & \textbf{A} &
\textbf{B} \\
\hline
1 & resonator, dc conn., Nb filters, bond pads      & 2500 & 2400 \\
2 & gate finger electrode in the transfer area      & 1800 & 1200 \\
3 & cross-linked PMMA gate insulator                  & 1600 &      \\
4 & central Au filter, connection resonator-filter  & 1600 &      \\
5 & contact electrodes in transfer area, Au filters & 420  & 1600 \\
6 & deep etching of the adjacent trenches           & 460  & 1700 \\
\hline
\end{tabular}
\end{center}
\caption{Coplanar waveguide resonator quality factors \Qcav\ of devices A and B
measured after subsequent lithographic fabrication steps.\label{tabq}
Some of the mentioned elements are only present in one device; see the text
for details. In particular, A contains only Au-based filters, while B contains 
only Nb-based filters.}
\end{table}

As first fabrication step (step 1 in Table \ref{tabq}) the coplanar waveguide
resonator, the dc bond pads, and their connecting lines are patterened in the
niobium film using optical lithography (blue layer in
Fig.~\ref{figDevices}(a-c)). Positive resist Microposit S1813 is spin-coated,
exposed with a mask-aligner, and developed using developer AZ300-47. For the
etch process, argon and sulphur hexafluorid is used in an Oxford Plasmalab
reactive ion etching (RIE) system.

Next, electron beam lithography is used to define the 100\,nm wide gate finger 
(red, in the center of Fig.~\ref{figDevices}(c); step 2 in Table \ref{tabq}). A 
polymethyl methacrylate (PMMA) bilayer (first layer molecular weight 200k 9\% 
in anisole, second layer 950k 2\% in anisole) is spin-coated, exposed, and 
developed with a mixture of methyl isobutyl ketone (MIBK) and isopropanole 
(IPA) in volume ratio 1:3. 10\,nm titanium as adhesive layer are 
sputter-deposited, followed by thermal evaporation of 50\,nm gold and lift-off 
in hot acetone.

In the following step, the gate insulator (brown in Fig.~\ref{figDevices}(a-c); 
step 3 in Table \ref{tabq}) is deposited as cross-linked PMMA: a PMMA
bilayer is spin-coated and locally overexposed by a factor $\sim$ 20. This
leads to cross-linking of the resist molecules and thus a $\sim$200\,nm thick 
insulator layer insoluble in acetone and other process solvents, with a low 
relative dielectric constant of typically\cite{book-brydson} 
$\epsilon_\text{PMMA}\simeq 3$.

Subsequently, the meander filter for the gate contact at the center of the 
resonator (red in Fig.~\ref{figDevices}(a), step 4 in Table \ref{tabq}) is 
defined, again via electron beam lithgraphy with a PMMA bilayer, metallization, 
and lift-off; here the gold thickness is 200\,nm. The gold meander has a strip 
width of 500\,nm and consist of 160 turns of each 18\,\textmu{}m length.

This is repeated for the four contact electrodes (source, drain, and two 
cutting elecrodes) in the transfer area and the meander filters connecting 
them (again, red in Fig.~\ref{figDevices}(a-c), step 5 in Table \ref{tabq}). 
The gold meanders for the four contacts have $\sim$ 120 turns of each 
35\,\textmu{}m length.

As last step, deep trenches are etched on both sides of the contact electrodes 
to allow insertion of the quartz forks during nanotube transfer (geometry not 
shown in the figure; step 6 in Table \ref{tabq}).\cite{forktransfer} We 
spin-coat two layers of AZ9260 photo resist and expose it using a mask aligner. 
AZ400K : water in ratio 1:2 is used as developer, follwed by RIE etching with 
argon and sulphur hexafluoride to a depth of $\sim$ 10\,\textmu{}m.

\section{Device geometry and fabrication steps, device B}

Device B, depicted in Fig.~\ref{figDevices}(d-e), is an improved version where
both the device geometry and (out of necessity) the fabrication steps have been
adapted. The most distinct change is the introduction of niobium-based 
T-filters in each dc connection,\cite{apl-hao-2014} as shown in
Fig.~\ref{figDevices}(e). These consist of a spiral inductor around a bond pad
with 2\,\textmu{}m thick turns separated by a 2\,\textmu{}m wide gap, an
interdigital capacitor that couples to the ground plane via 100 meshing fingers
on both sides, 117\,\textmu{}m long and 2\,\textmu{}m wide separated by gaps of
2\,\textmu{}m, and finally a meander inductor 2\,\textmu{}m wide, separated by
2\,\textmu{}m gaps, with 100 turns of each 198\,\textmu{}m length. For the
definition of the coplanar waveguide resonator a generic photomask without dc
contacts is used; the filters are added subsequently via electron beam
lithography and a second identical reactive ion etching process patterning the
niobium layer (i.e., included in step 1 in Table \ref{tabq}).

To reduce the capacitive coupling between gate finger and contact electrodes,
the contact electrodes (step 5) are additionally shortened; the ``coupling 
length'' $L$, where gate and contacts run in parallel, is reduced from about 
155\,\textmu{}m in device A to 86\,\textmu{}m in device B.

\section{Transmission measurements at 4.2K}

\begin{figure}[t]
\includegraphics{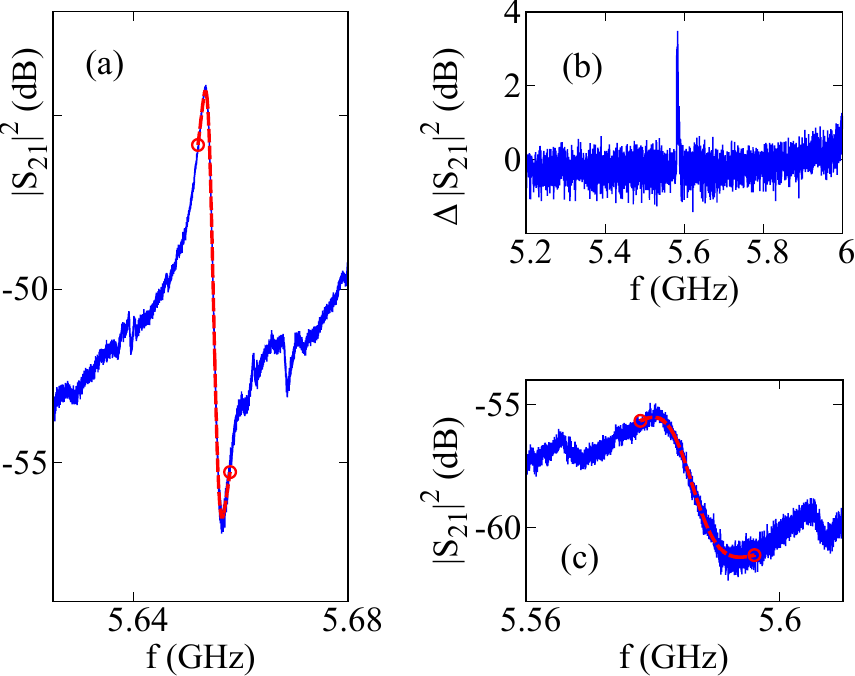}
\caption{Example VNA transmsission measurements at $T=4.2\un{K}$, with the
device immersed into liquid helium. (a) Device B, immediately after definition
of the coplanar waveguide resonator (step 1); $P=-20\un{dBm}$, raw
transmission data. (b,c) Device A, after all fabrication (step 6). (b)
Transmission difference of device A between measurements at VNA output power
$-20\un{dBm}$ and $+20\un{dBm}$. (c) Raw data plot of device A for 
$P=-20\un{dBm}$. In (a) and (c) the red line indicates the fit function as well 
as the fitted region for the evaluation of \Qcav.}
\label{figMeasurements}
\end{figure}

To identify detrimental processes, the devices were glued onto a sample 
carrier, bonded, and cooled down in a liquid helium vessel for microwave 
transmission measurement after each fabrication step. The temperature of the 
liquid helium, $T=4.2\un{K}$, is close enough to the critical temperature of 
niobium $T_\text{Nb}=9.2\un{K}$ to still lead to a reduction of intrinsic 
resonator quality factors,\cite{remo, toscres, pr-mattis-1958, jltp-gao-2008} 
however, any fabrication-induced reduction that is already visible here will 
also impact measurements at lower temperature. Thermalization of the device was 
done directly by immersion into the liquid helium, without any further 
low-temperature attenuation or isolation of the cables.

Figure~\ref{figMeasurements} displays example transmission measurements. 
The curve of Fig.~\ref{figMeasurements}(a) shows the transmission of device B 
directly after definition of the coplanar waveguide resonator (step 1), at a 
vector network analyzer (VNA) output power $P=-20\un{dBm}$. 
Fig.~\ref{figMeasurements}(b) shows a large frequency range plot of the 
difference in dB of two transmission measurements (i.e., the ratio of the 
measured transmissions) at $P=-20\un{dBm}$ and $P=+20\un{dBm}$, now for device 
A after all fabrication steps (step 6). This allows an easy identification of 
the resonance among a noisy background, since at larger incident power 
the superconductivity within the coplanar waveguide resonator breaks down first.
The corresponding raw data for device A, step 6 at $P=-20\un{dBm}$ is plotted 
in Fig.~\ref{figMeasurements}(c).

Fano resonances in transmission as visible in Fig.~\ref{figMeasurements} are a 
well-known phenomenon and caused by parasitic channels bypassing the resonator. 
They can be modeled with the expression\cite{apl-khalil-2012, jap-petersan-1998}
\begin{equation}\label{eq-s21fano}
S_{21}(f) =\frac{A}{1+2 i \Qcav\; (f-f_0)/f_0} + r e^{i\theta}, 
\end{equation}
where $A$ describes the overall transmission of the resonator, \Qcav\ and $f_0$ 
its quality factor and resonance frequency, and $r$ and $\theta$ the 
transmission amplitude and phase of the parasitic channel. For extracting the 
quality factors, we fit $\left| S_{21}(f) \right|^2$ to a selected interval of 
the measurement data. Due to the varying and irregular signal background, this 
selection clearly influences the result; this is the main source of error for 
the extracted \Qcav\ and the reason why only rounded values are given in 
Table~\ref{tabq}. In Fig.~\ref{figMeasurements}(a) and 
Fig.~\ref{figMeasurements}(c), the red line indicates the best fit function and 
the used interval.

The results are summarized in Table~\ref{tabq}. Both devices start out at 
$\Qcav \simeq 2500$. Device A retains $\Qcav \simeq 1600$ until after the 
deposition of the central bias connection to the resonator. With the 
fabrication of the source/drain electrodes, however, the quality factor sharply 
drops to $\Qcav \simeq 450$, a value quite close to the one observed in 
Refs.~\onlinecite{optomechanics, modelingomit}. The quality factor of device B,
initially similar as for device A, decreases with the deposition of the gate 
finger, but then remains near $\Qcav \simeq 1600$ until the end of the 
fabrication.

We can conclude that the optimizations have a clear effect; after all
lithograhic steps, device B with its large, niobium-based filters and the 
shorter ``coupling length'' $L$ of the electrodes, where they run parallel to
the gate finger, has a quality factor higher by approximately a factor 3.5 
compared to device A. In addition, since in device A the definition of the 
source/drain electrodes was the critical step reducing \Qcav, it is likely that 
the reduction of $L$ plays a role. From the experimental data it not 
possible yet to decide on the precise impact of the filter circuits.

\section{Comparison with numerical modeling}

To gain further insight into the nature of resonator losses and the 
effectiveness of the different types of filters, the software package Sonnet 
Professional\cite{sonnet} has been used to model a simplified circuit geometry. 
Sonnet uses the so-called method of moments\cite{book-harrington-1993, 
ieeemtt-rautio-1987} to calculate scattering matrix elements between circuit 
ports and is widely applied to problems in superconducting coplanar circuitry.

\begin{figure}[t]
\includegraphics{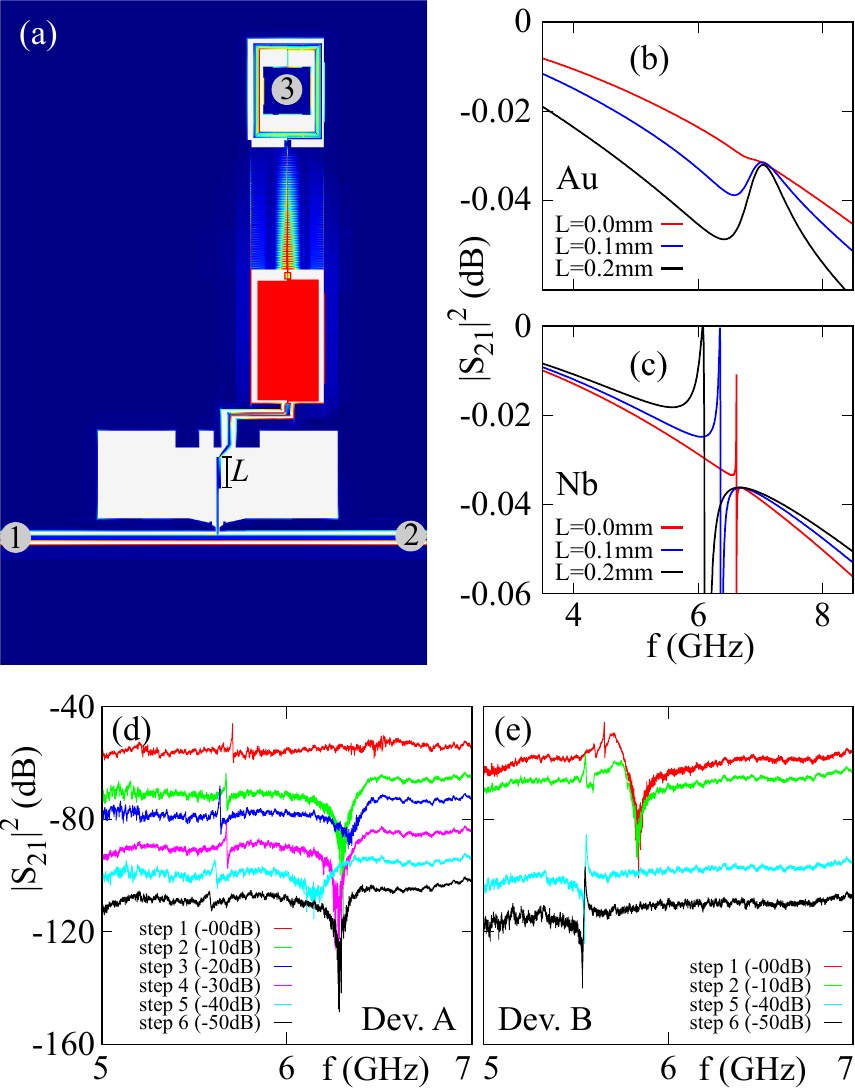} 
\caption{Numerical modeling of strongly simplified circuits using Sonnet 
Professional.\cite{sonnet} 
(a) Model geometry in the case of device B, including a segment of transmission 
line, a nanotube deposition area with closeby gate and contact electrodes ($L= 
0.1\un{mm}$), and a niobium T-filter with bond pad. Colors indicate the local 
current density at $f=6.38\un{GHz}$, the location of a filter resonance (red
corresponds to high current density). (b,c) Calculated power transmission
$\left|S_{21}\right|^2$ from port 1 to port 2, for (b) the Au-based filter
geometry (see text) and (c) the Nb-based filter geometry, in each case for
$L=0, 0.1, 0.2\un{mm}$. (d,e) Measured large frequency range transmission of
(d) device A and (e) device B, following each fabrication step. The curves have
been offset by 10\,dB each for visibility. Both the cavity resonance at
$f\simeq 5.6\un{GHz}$ and the filter resonance are visible.
}
\label{figSimulations}
\end{figure}

The substrate is modeled as stack of lossless insulators.
The circuit geometry is segmented by the software in the initial calculation
stages, with a minimum length scale set by user preference. To reduce 
calculation time we replace the fine gate finger electrode from the device 
geometries of Fig.~\ref{figDevices} as well as the contact electrodes with two 
5\,\textmu{}m wide niobium stripes running in parallel at a distance of
2\,\textmu{}m for a length of $L$, see Fig.~\ref{figSimulations}(a). In 
addition, for simplicity, the nanotube transfer region with contacts and 
filters is attached to a coplanar waveguide segment instead of a 
resonator.

Figure~\ref{figSimulations}(a) shows the actual model geometry used in 
correspondence to device B. Ports 1 (signal input) and 2 (signal output) are 
$Z=50\,\Omega$ terminated at the box walls of the calculated volume. A third 
port 3 connects the bond pad out of plane to ground; to approximate a dc wire 
connection, here we assume a termination with $R_3=0$ as well as an inductance 
of $L_3=2\,\text{nH}$, the rule-of-thumb value for a 2\,mm long wire bond.

In the variant of the geometry used for approximating device A, the niobium 
filter is replaced by a straight connection to the bond pad, and a fine gold 
meander identical to the one in the lithography drawing of device A is 
introduced directly at the edge of the transfer area.

Figures~\ref{figSimulations}(b) and \ref{figSimulations}(c) display the 
calculation result for the Au-based filter and the Nb-based filter geometry, 
respectively. Both geometries lead to a clear filter resonance in the region 
$6\un{GHz} \le f_r \le 7\un{GHz}$. The current density plotted in 
Fig.~\ref{figSimulations}(a) illustrates this at $f=6.38\un{GHz}$ and 
$L=0.1\un{mm}$. For the resistive gold meander, Fig.~\ref{figSimulations}(b), a 
strongly broadened Fano function results, with an initially decreasing 
transmission at low frequency as well as an overall decrease of transmission.
Measurement of the transmission of device A over a large frequency range, see 
Fig.~\ref{figSimulations}(d), indeed confirm the presence of such a filter
resonance with the same behaviour as soon as the gate finger electrode has been 
deposited and thus allows for a coupling between cavity and dc contact 
connections.

The filter resonance of the larger, niobium based filter is considerably 
sharper due to the superconductivity of the material (modelled as lossless 
metal), see Fig.~\ref{figSimulations}(c). In addition, the Fano behaviour 
displays different polarity: the filter initially leads for frequencies below 
its resonance to suppressed damping, via constructive interference of the 
reflected signal. For larger $L$, the resonance moves to lower frequencies.
Fig.~\ref{figSimulations}(e), a measurement of device B, clearly agrees with 
this result for the first two fabrication steps.

The filter resonance appears to be absent after deposition of the source/drain 
contacts (step 5). While from the data no definite explanation for this can be 
given, a striking detail is that here also the Fano shape of the {\em 
microwave cavity} resonance has changed its polarity. Via the much stronger 
coupling of the cavity to the dc connection and its filter, the filter 
resonance may have moved to significantly lower frequency. We can speculate 
that it may have merged with the cavity resonance or passed it entirely, 
with both signal contributions phase-shifted and again interfering 
constructively.

\begin{figure}[t]
\includegraphics{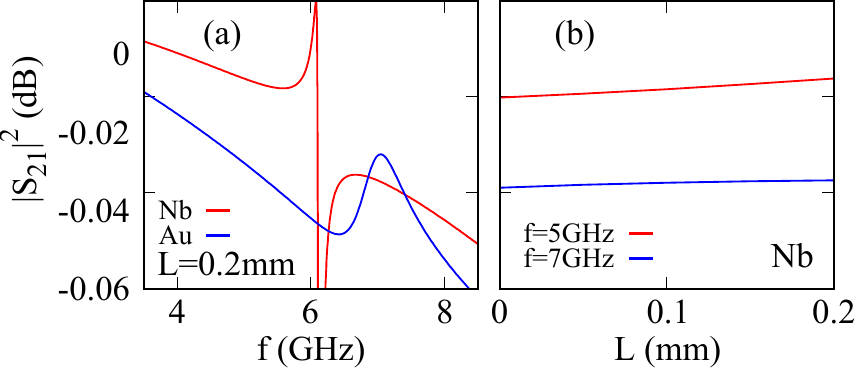} 
\caption{
(a) Model power transmission $\left|S_{21}\right|^2$ as function of frequency 
for the two filter types, at $L=0.2\un{mm}$. 
(b) Model power transmission $\left|S_{21}\right|^2$ as function of coupling 
length $L$ for the Nb-based filter, at $f=5\un{GHz}$ and $f=7\un{GHz}$. 
}
\label{figSim2}
\end{figure}

Additional modeling results are shown in Fig.~\ref{figSim2}. The power 
transmission $\left|S_{21}\right|^2$ as function of frequency for the two 
filter types is compared in Fig.~\ref{figSim2}(a) at $L=0.2\un{mm}$. Again, 
this plot clearly demonstrates that the Au-based filters lead to damping, while 
at proper choice of parameters the Nb-based filter reflects the GHz signal back 
into the circuit for constructive interference. Figure~\ref{figSim2}(b) plots 
the transmission for the Nb-based geometry as function of the coupling length 
$L$ for $f=5\un{GHz}$ and $f=7\un{GHz}$, i.e., below and above the filter 
resonances. As expected for the reflection mechanism, a stronger coupling 
of the dc connections here even counterintuitively improves the signal.

\section{Conclusions and outlook}

Two coplanar waveguide resonator devices including dc electrodes for microwave
optomechanical experiments involving carbon nanotubes have been fabricated, 
with tests of low temperature GHz transmission properties after each step.
Device A, identical in geometry with the device of 
Refs.~\onlinecite{optomechanics, modelingomit}, includes a resistive gold 
meander in the dc connections, device B a niobium-based T-filter. We show 
experimentally that the quality factor of device A sharply drops with the 
fabrication of the source/drain electrodes, while device B retains a 
significantly higher \Qcav\ up to the end of processing.

Model calculations using Sonnet Professional\cite{sonnet} on a simplified 
device geometry confirm these results. While the gold-based filter geometry 
leads to damping of the reflected signal, the niobium-based geometry can better 
reﬂect the signal back into the circuit, for constructive interference, 
minimizing losses and thereby in a full device maximizing \Qcav. For the latter 
effect, a filter resonance plays an important role, a fact that will have to be 
taken into account for future device geometry planning.

While for superconducting two-level systems already multiple mechanisms 
to enhance optomechanical coupling have been proposed and 
implemented,\cite{ncomms-pirkkalainen-2015, cphys-schmidt-2020} for the 
particularly interesting case of 
carbon nanotube resonators this research is still at the start. Aside from 
quantum capacitance effects\cite{optomechanics, modelingomit, 
prb-manninen-2022}, also integrating carbon nanotubes as variable Josephson 
inductors is expected to lead to a strong coupling
amplification.\cite{prl-heikkila-2014, njp-rimberg-2014} In addition to a 
modified GHz circuit geometry, this requires transparent contacts between 
superconductor and carbon nanotube,\cite{prl-buitelaar-2003, 
nature-jarillo-2006, prl-jorgensen-2006} in a device that keeps the carbon 
nanotubes suspended and thus nanomechanically active. Recent 
results\cite{nres-kaikkonen-2020} with gate voltage dependent critical 
currents in suspended, as-grown single-wall carbon nanotubes of up to 53\,nA 
indicate that molybden-rhenium alloys\cite{srep-schneider-2012, 
remo} provide a solution to this challenge, showing the way 
towards future optomechanical hybrid devices.

\section*{Acknowledgments}

The authors acknowledge funding by the Deutsche Forschungsgemeinschaft via
grants Hu 1808/4 (project id 438638106) and Hu 1808/5 (project id 438640202).
This work was supported by the Academy of Finland project 312295 (CoE, Quantum 
Technology Finland).
A.~K.~H. acknowledges support from the Visiting Professor program of 
the Aalto University School of Science. We would like to thank O.~Vavra for 
experimental help, Ch.~Strunk and D.~Weiss for the use of experimental 
facilities, and A.~N.~Loh for insightful discussions. The measurement data has 
been recorded using Lab::Measure\-ment \cite{labmeasurement}.

\section*{References}

\bibliography{paper}

\end{document}